\documentclass[pra,11pt, twoside, nofootinbib]{revtex4}

\usepackage{titletoc}
\usepackage{hyperref}

\usepackage{tabularx}
   \newcolumntype{C}{>{\centering\arraybackslash}X}
   \newcolumntype{L}{>{\raggedright\arraybackslash}X}
   \newcolumntype{R}{>{\raggedleft\arraybackslash}X}
  
\usepackage{amsmath}
\usepackage{amsfonts}
\usepackage{amssymb}
\usepackage{amstext}
\usepackage[sort&compress]{natbib}
\usepackage{amsthm}
\usepackage{latexsym}
\usepackage{graphicx}
\usepackage{textcomp}
\usepackage{color}
\usepackage{mathtools}
\usepackage {diagbox}
\usepackage{bm}
\usepackage{multirow}

\makeatletter
\newcommand*{\rom}[1]{\expandafter\@slowromancap\romannumeral #1@}
\makeatother

\newtheorem{Theorem}{Theorem}

\theoremstyle{definition}

\newcolumntype{L}[1]{>{\raggedright\arraybackslash}p{#1}}
\newcolumntype{C}[1]{>{\centering\arraybackslash}p{#1}}
\newcolumntype{R}[1]{>{\raggedleft\arraybackslash}p{#1}}

\newcommand{\be}{\begin{equation}}
\newcommand{\ee}{\end{equation}}
\newcommand{\bea}{\begin{eqnarray}}
\newcommand{\eea}{\end{eqnarray}}

\newcommand{\ket}[1]{|#1\rangle} 
\newcommand{\bra}[1]{\langle#1|} 
\newcommand{\ketbra}[2]{|#1\rangle\langle#2|} 

\DeclareMathOperator{\tr}{Tr}

\newcommand{\cH}{{\mathcal H}}

\newcommand{\cK}{{\cal K}}

\newcommand{\cU}{\mathcal{U}}

\def\>{{\rangle}}
\def\<{{\langle}}

\def\lsim{\mathrel{\rlap{\lower4pt\hbox{\hskip1pt$\sim$}}
		\raise1pt\hbox{$<$}}}                
\def\gsim{\mathrel{\rlap{\lower4pt\hbox{\hskip1pt$\sim$}}
		\raise1pt\hbox{$>$}}}                

\def\app#1#2{%
  \mathrel{%
    \setbox0=\hbox{$#1\sim$}%
    \setbox2=\hbox{%
      \rlap{\hbox{$#1\propto$}}%
      \lower1.1\ht0\box0%
    }%
    \raise0.25\ht2\box2%
  }%
}

\begin{document}
	
\title{Generic Entanglement Entropy for Quantum States with Symmetry}
	
\author{Yoshifumi Nakata$^{1,2}$} \footnote{nakata@qi.t.u-tokyo.ac.jp}
\author{Mio Murao$^3$}
\affiliation{
$^1$Photon Science Center, Graduate School of Engineering,~The University of Tokyo, Bunkyo-ku,~Tokyo 113-8656, Japan\\
$^2$JST, PRESTO, 4-1-8 Honcho, Kawaguchi, Saitama, 332-0012, Japan\\
$^3$ Department of Physics, Graduate School of Science, The University of Tokyo, 7-3-1, Hongo, Bunkyo-ku, Tokyo 113-0033, Japan
}
	
\begin{abstract}
When a quantum pure state is drawn uniformly at random from a Hilbert space, the state is typically highly entangled. This property of a random state is known as generic entanglement of quantum states and has been long investigated from many perspectives, ranging from the black hole science to quantum information science. In this paper, we address the question of how symmetry of quantum states changes the properties of generic entanglement. More specifically, we study bipartite entanglement entropy of a quantum state that is drawn uniformly at random from an invariant subspace of a given symmetry. We first extend the well-known concentration formula to the one applicable to any subspace and then show that 1. quantum states in the subspaces associated with an axial symmetry are still highly entangled, though it is less than that of the quantum states without symmetry, 2. quantum states associated with the permutation symmetry are significantly less entangled, and 3. quantum states with translation symmetry are as entangled as the generic one. We also numerically investigate the phase-transition behavior of the distribution of generic entanglement, which indicates that the phase transition seems to still exist even when random states have symmetry.
\end{abstract}

\maketitle

\section{Introduction}

Randomness is often an important resource in information processing. This is true even in the quantum regime, where quantum randomness is often represented by a \emph{random state}, a~quantum pure state that is drawn uniformly at random from a Hilbert space. A~random state is known to be extremely useful and is used in numerous quantum information protocols, from~communication~\cite{DLT2002,BJ2012,SDTR2013} and computation tasks~\cite{BH2013,BFNV2018,GOOGLE2019} to benchmarking quantum devices~\cite{KLRetc2008,MGE2012,RLL2009,CGJetc2013}. The~origin of its usefulness can be traced back to the counter-intuitive property of a random state that it is typically extremely highly~entangled.

Quantum randomness is also the key to understanding the physics in complex quantum many-body systems. In~the last decade, based on various measures of quantum randomness, such~as scrambling~\cite{SS2008,LSHOH2013}, operator entanglement~\cite{BL2005,HQRY2016}, and~out-of-time-ordered correlators~\cite{RY2017}, quantum~randomness in various complex quantum many-body systems has been intensely studied. It eventually turns out that randomness is indeed the key to connect the physics in quantum chaotic systems and that of quantum black hole, revealing an exotic relation between them~\cite{K2015,K2015Feb,SS2015}. The~measures used in the analyses are all to elucidate entanglement of a random state. Hence, entanglement of a random state plays a crucial role in the~approach.

The entanglement of a random state is often called \emph{generic} entanglement of quantum states due to the fact that a random state is uniformly distributed in a Hilbert space and can be considered to represent generic properties of quantum pure states. Generic entanglement has been especially studied in terms of a bi-partition of the system. It was first pointed out that in terms of the entanglement entropy, generic entanglement in a large system is typically extremely concentrated around a nearly, but~not exactly, maximum value~\cite{L1978,Page1993,FK1994,S-R1995,S1996}.
The analysis was then extended to a probabilistic statement~\cite{HLW2006}, revealing the relation with quantum statistical mechanics~\cite{PSW2006}, and~to the higher moments of the distribution of entanglement entropy~\cite{G2007,FMPPS2008,PFPPS2010,NMV2010,PFGPPS2011,NMV2011,FFPPY2013,FPPSY2019}.
In~particular, higher moments were studied in great detail using the technique of random matrix theory. It was shown that the probability density function of the distribution has two singularities, splitting the distribution into three different \emph{entanglement phases} with different entanglement spectra. Since the entanglement spectrum characterizes topological orders of the state, this implies that there exists yet another intriguing relation between a random state and an exotic quantum many-body~physics.

{There is also a close relation between generic entanglement and quantum error correction, one of the key concepts in quantum information science. It is well-known that a randomly chosen unitary is typically a good encoder of quantum information~\cite{HOW2005,HOW2007,ADHW2009}. It is recently pointed out that a certain property of generic entanglement is responsible for this~\cite{NWK2020}. Hence, revealing the properties of generic entanglement will help our understanding of why quantum error correction works well, even~providing real applications of generic entanglement in quantum information science.}

Most of these studies of generic entanglement focus on the random state  that is uniformly distributed over the \emph{whole} Hilbert space. However, quantum many-body systems often have symmetry, restricting the distribution of states into that over the invariant subspace of the symmetry. Hence, the~aforementioned results about generic entanglement cannot be directly applied to complex quantum many-body systems with symmetry. It is also worth pointing out that symmetry is the guiding principle in many-body physics, allowing us to understand intriguing many-body phenomena, such~as thermal and quantum phase transitions, in~a unified manner. Thus, it will be interesting to take symmetry into account in the study of generic~entanglement.

In this paper, we address the question of how symmetry of quantum systems changes the properties of generic entanglement. We specifically investigate bipartite entanglement of random states in invariant subspaces.
To this end, we first provide a general formula that is useful to analyze the distribution of entanglement over a random state in any subspace. We then apply the formula to investigate the generic entanglement in invariant subspaces associated with a given symmetry. We especially consider three symmetries, 1. axial symmetry that leads to the conservation law of a component of angular momentum, 2. permutation symmetry that characterizes indistinguishable bosons and fermions, and~3. translation symmetry that defines the structure of a lattice. 
{We particularly focus on these symmetries since the axial symmetry is the one used in Ref.~\cite{NWK2020} that pointed out the relation between generic entanglement and quantum error correction, the~permutation symmetry is commonly believed to result in weak entanglement, and~the translation symmetry is important in relation to the area law of entanglement. We however note that the formula we derive can be applied to any symmetries.}
We then find that compared to generic entanglement of a random state without symmetry, the~axial and permutation symmetries reduce the amount of entanglement by a constant and a significant degree, respectively, while the translation symmetry does not lead to a significant reduction. 
We also numerically study whether the distribution of entanglement over random states in invariant subspaces has phase-transition-like behaviors. Although~it is less conclusive due to a large finite-size effect, we show that certain entanglement phases seem to exist even when a random state has permutation or translation~symmetry.

This paper is organized as follows. 
We start with preliminaries in Section~\ref{Sec:Prel} and overview properties of generic entanglement in Section~\ref{S:NoSym}. In~Section~\ref{S:Concentration}, our main technical tool in the analysis is provided.
We then investigate generic entanglement with axial, permutation, and~translation symmetries in Sections~\ref{S:AxialSymmetry}--\ref{S:TransSym}, respectively.
We finally numerically analyze possible entanglement phases of random states with symmetry in Section~\ref{S:EntPhase}. After~we make a remark on the feasibility of generic entanglement in Section~\ref{S:F}, we conclude with a summary and discussions in Section~\ref{S:Conclusion}.

\section{Preliminaries}  \label{Sec:Prel}

Throughout the paper, we consider a quantum system $\Lambda$ composed of $n$ qudits, whose Hilbert space is $(\mathbb{C}^{d})^{\otimes n}$, and~its bi-partition into subsystems $A$ and $\bar{A}$, which consist of with $n_A$ and $n_{\bar{A}}$ qudits, respectively. We assume that $n_A \leq n_{\bar{A}}$.
For Hilbert spaces, and~operators, we often write the systems on which they are defined in the superscript. For~instance, $\mathcal{H}^{\Lambda}$ is the Hilbert space associated with the system $\Lambda$, and~$X^{A}$ is an operator $X$ acting on the system $A$. A~reduced operator on $A$ of $\rho^{\Lambda}$ is denoted simply by $\rho^A$, i.e.,~$\rho^A = \tr_{\bar{A}} [\rho^{\Lambda}]$. We denote by $I$ the identity operator, and~by $\Pi$ projection onto some~subspace.

\subsection{Haar Measure, Haar Random Unitaries, and~Haar Random~States}
On a unitary group with finite degree, there exists the unique unitarily invariant probability measure, known as the \emph{Haar measure}. We denote it by ${\sf H}$, which satisfies the following properties: for~any subset $\mathcal{V}$ of unitaries and for any unitary $U$,
\begin{equation}
{\sf H}(\mathcal{V}U) = {\sf H}(U\mathcal{V}) = {\sf H}(\mathcal{V}) \geq 0, \text{\ and \ } \int {\sf H}(U) dU = 1.
\end{equation}
The integral is taken over the whole unitary group.
When a unitary $U$ is chosen from the unitary group uniformly at random with respect to the Haar measure ${\sf H}$, we denote it by $U \sim {\sf H}$ and call it a \emph{Haar random unitary}.

Let $U \sim {\sf H}$ be a Haar random unitary acting on a Hilbert space $\mathcal{H}$. The~state $\ket{\phi}$ obtained by applying $U$ to a fixed canonical pure state $\ket{\phi_0} \in \mathcal{H}$ is called a \emph{Haar random state}. 
With a slight abuse of notation, we denote a Haar random state as $\ket{\phi} \sim {\sf H}$.
Due to the uniform distribution of a Haar random unitary, the~distribution of a Haar random state does not depend on the choice of the canonical state $\ket{\phi_0}$ and is uniform in the Hilbert space $\cH$. Thus, it is often used to study generic properties of quantum pure states.
In this paper, we often use the Haar measure on the unitary group acting on a subspace $\cK$ of a Hilbert space $\cH$. The~Haar measure on the unitary group acting only on the subspace $\cK$ is denoted by ${\sf H}_{\cK}$.

Since the Haar measure is a probability measure, we can think of an average of a function $f(\ket{\phi})$ of a state $\ket{\phi}$ over the Haar measure ${\sf H}$. We denote the average by $\mathbb{E}_{\ket{\phi} \sim {\sf H}}[f(\ket{\phi})]$.
Similarly, the~probability with respect to the Haar measure is denoted by ${\rm Prob}_{\ket{\phi} \sim {\sf H}}$.

\subsection{Entanglement Entropy, and~Entanglement~Spectrum}
For a pure state $\ket{\phi}$ in $\mathcal{H}^{\Lambda}$, we quantify the amount of entanglement with respect to the bi-partition $A$ and $\bar{A}$ by the von Neumann entropy of the reduced density matrix $\phi^A$ in $A$. That is, we use
\begin{equation}
E_A(\ket{\phi}) := S(\phi^A),
\end{equation}
as a measure of bi-partite entanglement of $\ket{\phi}$, where $S(\rho) := -\tr[ \rho \log \rho]$ is the von Neumann entropy. The~measure $E_A(\ket{\phi})$ is often referred to as the \emph{entanglement entropy} of $\ket{\phi}$ and takes the value between $0$ for separable states and $n_A \log d$ for maximally entangled~states.

For a given pure state $\ket{\phi} \in \cH^{\Lambda}$, the~distribution of the eigenvalues of the reduced density matrix $\phi^A$ in decreasing order is called an \emph{entanglement spectrum} of $\ket{\phi}$ in $A$.

\section{Generic Entanglement without~Symmetry} \label{S:NoSym}

It is well-known that a random state in an $n$-qudit system $\Lambda = A \bar{A}$ is typically extremely highly entangled between $A$ and $\bar{A}$, which has been extensively studied in the literature~\cite{L1978,Page1993,FK1994,S-R1995,S1996,HLW2006,G2007,FMPPS2008,PFPPS2010,NMV2010,PFGPPS2011,NMV2011,FFPPY2013,FPPSY2019}. For~instance, the~average entanglement entropy of a Haar random state satisfies~\cite{L1978,Page1993,FK1994,S-R1995,S1996,HLW2006}
\begin{equation}
\mathbb{E}_{\ket{\phi} \sim {\sf H}} [E_A(\ket{\phi})] > n_A \log d - \frac{d^{-n+2n_A}}{2 \ln 2}.
\end{equation}
Since the maximum value of the entanglement entropy is $n_A \log d$, this implies that the average is exponentially close to the maximum when $1 \ll n_A$.
This statement was later strengthened to the probabilistic statement that the entanglement entropy of a random state strongly concentrates around its average~\cite{HLW2006}.

\begin{Theorem}[Theorem III.3 in Ref.~\cite{HLW2006}] \label{Thm:TypEnt}
Let $\Lambda$ be a composite system, $\Lambda=A\bar{A}$, and~$n$ and $n_A$ be the number of qudits in $\Lambda$ and $A$, respectively, that satisfy $n/2 \geq n_A \geq \log 3/\log d$.
For a random state $\ket{\phi} \sim {\sf H}$ in $\mathcal{H}^{\Lambda}$, it holds that $\forall \epsilon >0$,
\begin{equation}
{\rm Prob}_{\ket{\phi} \sim {\sf H}} \biggl[ E_A(\ket{\phi}) \geq n_A \log d  - \frac{d^{-n+2 n_A}}{\ln 2} - \epsilon \biggr] > 1- \exp\biggl[ -\frac{(d^n - 1) \epsilon^2}{8 \pi^2 \ln 2 (n_A \log d)^2} \biggr]. \label{Eq:4}
\end{equation}
\end{Theorem}
Since the probability is close to $1$ doubly exponentially in the number $n$ of qudits in $\Lambda$, this clearly shows that it is extremely unlikely that the entanglement entropy of a random state takes the value far from its~average.

Theorem~\ref{Thm:TypEnt} is not only of theoretical interest, but~also has implications onto many topics in quantum physics. In~particular, in~the context of the condensed-matter physics, where qudits are often aligned on a lattice, a~pure state is said to obey the \emph{volume law of entanglement} when the entanglement entropy of the state is proportional to the number of qudits in the subsystem. Although~the volume law does not hold in most many-body systems, it is expected to hold when the dynamics of the system is sufficiently scrambling, which is likely to be the key feature bridging quantum chaos and quantum gravity. Hence, the~volume law of entanglement is considered to be one of the diagnostic features of complex many-body quantum systems.
In this context, Theorem~\ref{Thm:TypEnt}, stating that a state generated by a random unitary dynamics typically obeys the volume law of entanglement, implies that typical unitary dynamics without any restriction should be highly~chaotic.

The entanglement entropy of a Haar random state also has an intriguing property, namely~`phase~transitions' of the distribution~\cite{G2007,FMPPS2008,PFPPS2010,NMV2010,PFGPPS2011,NMV2011,FFPPY2013,FPPSY2019}. This was first studied based on the purity of reduced density matrices, and~was then extended to the R\'{e}nyi entropies and eventually to the von Neumann entropy, i.e.,~the entanglement entropy. The~probability density function of the entanglement entropy $E_A(\ket{\phi})$ over a Haar random state $\ket{\phi} \sim {\sf H}$ has two singularities when $n_A \rightarrow \infty$. Thus, the~distribution of the entanglement entropy is split into three regimes, which are sometimes called \emph{separable}, \emph{typical}, and~\emph{maximally entangled} phases. Each entanglement phase has a different characteristic entanglement spectrum. Thus, although~the average of entanglement entropy of a Haar random state is nearly maximum, its distribution has a rather rich~structure.

The main question in this paper is how symmetry of quantum states affects these properties of generic entanglement. This is of crucial importance when we are interested in the implications of generic entanglement on the physics in complex quantum many-body systems with~symmetry.

\section{Concentration of Entanglement Entropy of a Random State in a~Subspace} \label{S:Concentration}

To investigate the entanglement entropy of a random state with symmetry, we use the same technical tool as used to show Theorem~\ref{Thm:TypEnt}, which is the so-called \emph{concentration phenomena} of the Haar measure~\cite{L2001}. It states that any real-valued function of a Haar random state strongly concentrates around its average if the function is sufficiently smooth.
As the entanglement entropy is a real-valued function, it can be directly applied to the question we are interested in, leading to the following~Theorem.

\begin{Theorem} \label{Thm:ConEnt}
Let $\mathcal{H}^{\Lambda}$ be the Hilbert space of an $n$-qudit system $\Lambda = A\bar{A}$, and~$\mathcal{K} \subset \mathcal{H}^{\Lambda}$ be any $D_{\mathcal{K}}$-dimensional subspace. Let $\Omega_{\mathcal{K}}^A$ and $\Omega_{\mathcal{K}}^{\bar{A}}$ be a state on the subsystem $A$ and $\bar{A}$, defined by
\begin{equation}
\Omega_{\mathcal{K}}^A
:= 
\tr_{\bar{A}} \biggl[ \frac{\Pi^{\Lambda}_{\mathcal{K}}}{D_{\mathcal{K}}} \biggr],
\text{\ \  and\ \ }
\Omega_{\mathcal{K}}^{\bar A}
:= 
\tr_{A} \biggl[ \frac{\Pi^{\Lambda}_{\mathcal{K}}}{D_{\mathcal{K}}} \biggr],
\end{equation}
respectively, where $\Pi^{\Lambda}_{\mathcal{K}}$ is the projection onto $\mathcal{K}$. Then, for~a random state $\ket{\phi} \sim {\sf H}_{\mathcal{K}}$ in $\mathcal{K}$, and~$\forall \epsilon>0$, it holds that
\begin{equation}
{\rm Prob}_{\ket{\phi} \sim {\sf H}_{\mathcal{K}}} \bigl[ E_A(\ket{\phi})  \geq \bar{S}(\mathcal{K}) - \epsilon \bigl]
>
1-  \exp \biggl[- \frac{(D_{\mathcal{K}} +1) \epsilon^2}{72 \pi^3 \ln 2 (\ln R_{\mathcal{K}})^2} \biggr], \label{Eq:6}
\end{equation}
where
\begin{equation}
\bar{S}(\mathcal{K}) := -\log \bigl[ \tr (\Omega_{\mathcal{K}}^A)^2 + \tr (\Omega^{\bar{A}}_{\mathcal{K}})^2 \bigr] - \log \bigl[ 1- \frac{1}{D_{\mathcal{K}}+1} \bigr],
\end{equation}
and $R_{\mathcal{K}} = \max_{\ket{\phi} \in \mathcal{K}}[ {\rm supp} (\phi^A) ]$.
\end{Theorem}

Theorem~\ref{Thm:ConEnt} is a slight generalization of Theorem~\ref{Thm:TypEnt}, so that it is applicable to any subspace $\cK \subset \cH^{\Lambda}$. In~the case of $\cK = \cH^{\Lambda}$, Theorem~\ref{Thm:ConEnt} nearly recovers Theorem~\ref{Thm:TypEnt} except that the probability in Equation~\eqref{Eq:6} is worse than that in Equation~\eqref{Eq:4}. This is because the latter probability is obtained by using the median rather than the average. Using the same technique, it will be possible to slightly improve Equation~\eqref{Eq:6}. 

\begin{proof}[Proof of Theorem~\ref{Thm:ConEnt}]
The proof is based on Levy's lemma~\cite{L2001}. We particularly use the lemma in the form given in Ref.~\cite{HLW2006}, which is tailored to the entanglement entropy: for any $\epsilon > 0$, it holds that
\begin{equation}
{\rm Prob}_{\ket{\phi} \sim {\sf H}_{\mathcal{K}}}\bigl[ E_A(\ket{\phi}) \geq \mathbb{E}E_A - \epsilon \bigr] 
>
1-  \exp \biggl[- \frac{(D_{\mathcal{K}} +1) \epsilon^2}{72 \pi^3 \ln 2 (\ln R)^2} \biggr], \label{Ineq:Levy}
\end{equation}
where $\mathbb{E}E_A := \mathbb{E}_{\ket{\phi} \sim {\sf H}_{\mathcal{K}}}[E_A(\ket{\phi})]$ is the average of $E_A$ over the probability measure ${\sf H}_{\mathcal{K}}$ on $\mathcal{K}$. In~the following, we show that $\mathbb{E}_{\ket{\phi} \sim {\sf H}_{\mathcal{K}}}[E_A(\ket{\phi})] \geq \bar{S}(\cK)$.

We first use the monotonicity of the R\'{e}nyi entropy, i.e.,~$S(\rho) \geq -\log[ \tr \rho^2]$. Further using the Jensen's inequality, we obtain
\begin{equation}
\mathbb{E}E_A(\ket{\phi}) \geq - \log \bigl[ \mathbb{E} \tr[(\phi^A)^2] \bigr].
\end{equation}
We then introduce a system $X'$ of $X$ for $X = \Lambda, A, \bar{A}$, whose Hilbert space $\mathcal{H}^{X'}$ is isomorphic to $\mathcal{H}^X$, and~denote by $\mathbb{I}^{XX'}$ and $\mathbb{F}^{XX'}$ the identity and the swap operator on $XX'$, respectively. They are explicitly given by
\begin{equation}
\mathbb{F}^{XX'} = \sum_{i,j = 1}^{\dim \cH^X} \ketbra{b_i}{b_j}^X \otimes \ketbra{b_j}{b_i}^{X'}, \text{\ \ and\ \ }\mathbb{I}^{XX'} = I^X \otimes I^{X'},
\end{equation}
where $\{ \ket{b_i} \}$ is an orthonormal basis in $\mathcal{H}^X$. Note that the definition of $\mathbb{F}^{XX'}$ does not depend on the choice of the basis.
Using these operators and the so-called swap trick, i.e.,~$\tr [P^X Q^X] = \tr[(P^X \otimes Q^{X'})\mathbb{F}^{XX'} ]$ for any operator $P$ and $Q$ on $X$, it follows that
\begin{align}
\mathbb{E} \tr[(\phi^A)^2] &= \mathbb{E} \tr[(\phi^A \otimes \phi^{A'}) \mathbb{F}^{AA'}],\\
&= \mathbb{E} \tr[(\phi^{\Lambda} \otimes \phi^{\Lambda'}) ( \mathbb{F}^{AA'} \otimes \mathbb{I}^{\bar{A}\bar{A}'})],\\
&=  \tr \bigl[ \mathbb{E}[\phi^{\Lambda} \otimes \phi^{\Lambda'}] ( \mathbb{F}^{AA'} \otimes \mathbb{I}^{\bar{A}\bar{A}'}) \bigr].
\end{align}
Hence, it suffices to compute $\mathbb{E}_{\ket{\phi} \sim {\sf H}_{\mathcal{K}}}[\phi^{\Lambda} \otimes \phi^{\Lambda'}]$, which can be explicitly done using the unitary invariance of the Haar~measure.

For any unitary $U^{\mathcal{K}}$ acting on the subspace $\mathcal{K}$, it holds that
\begin{equation}
(U^{\mathcal{K}} \otimes U^{\mathcal{K}'} )
\mathbb{E}_{\ket{\phi} \sim {\sf H}_{\mathcal{K}}}[\phi^{\Lambda} \otimes \phi^{\Lambda'}]
(U^{\mathcal{K}} \otimes U^{\mathcal{K}'} )^{\dagger}
=
\mathbb{E}_{\ket{\phi} \sim {\sf H}_{\mathcal{K}}}[\phi^{\Lambda} \otimes \phi^{\Lambda'}].
\end{equation}
Due to the Schur-Weyl duality, this implies that $\mathbb{E}_{\ket{\phi} \sim {\sf H}_{\mathcal{K}}}[\phi^{\Lambda} \otimes \phi^{\Lambda'}]$ is given by a linear combination of the unitary representations of permutations between $\mathcal{K}$ and $\mathcal{K}'$, or~equivalently, a~linear combination of $\mathbb{I}^{\mathcal{K}\mathcal{K}'}$ and $\mathbb{F}^{\mathcal{K}\mathcal{K}'}$. In~terms of the operators defined on $\Lambda$, they are respectively given by
\begin{equation}
\mathbb{I}^{\mathcal{K}\mathcal{K}'} = \Pi_{\mathcal{K}}^{\Lambda} \otimes \Pi^{\Lambda}_{\mathcal{K}'}, \text{\ \ and \ \ }
\mathbb{F}^{\mathcal{K}\mathcal{K}'} = (\Pi_{\mathcal{K}}^{\Lambda} \otimes \Pi^{\Lambda}_{\mathcal{K}'}) \mathbb{F}^{\Lambda \Lambda'} (\Pi_{\mathcal{K}}^{\Lambda} \otimes \Pi^{\Lambda}_{\mathcal{K}'}).
\end{equation}
We now have $\mathbb{E}_{\ket{\phi} \sim {\sf H}_{\mathcal{K}}}[\phi^{\Lambda} \otimes \phi^{\Lambda'}] = \alpha \mathbb{I}^{\mathcal{K}\mathcal{K}'}  + \beta \mathbb{F}^{\mathcal{K}\mathcal{K}'}$ for some coefficients $\alpha$ and $\beta$. The~coefficients are determined from the conditions that
\begin{align}
&\tr \bigl[ \mathbb{E}_{\ket{\phi} \sim {\sf H}_{\mathcal{K}}}[\phi^{\Lambda} \otimes \phi^{\Lambda'}] \bigr]= 1, \\
&\tr \bigl[ \mathbb{E}_{\ket{\phi} \sim {\sf H}_{\mathcal{K}}}[\phi^{\Lambda} \otimes \phi^{\Lambda'}] \mathbb{F}^{\Lambda \Lambda'} \bigr] = 1.
\end{align}
Noting that $\tr \mathbb{I}^{\mathcal{K}\mathcal{K}'}  = D_{\mathcal{K}}^2$ and $\tr \mathbb{F}^{\mathcal{K}\mathcal{K}'}  = D_{\mathcal{K}}$, we obtain $\alpha = \beta = (D_{\mathcal{K}} (D_{\mathcal{K}}+1))^{-1}$ and so,
\begin{equation}
\mathbb{E}_{\ket{\phi} \sim {\sf H}_{\mathcal{K}}}[\phi^{\Lambda} \otimes \phi^{\Lambda'}]
=
\frac{\mathbb{I}^{\mathcal{K}\mathcal{K}'} + \mathbb{F}^{\mathcal{K}\mathcal{K}'}}{D_{\mathcal{K}} (D_{\mathcal{K}}+1)}.
\end{equation}
We thus arrive at
\begin{equation}
\mathbb{E} \tr[(\phi^A)^2] 
=  \frac{1}{D_{\mathcal{K}} (D_{\mathcal{K}}+1)}
\tr[ (\mathbb{I}^{\mathcal{K}\mathcal{K}'} + \mathbb{F}^{\mathcal{K}\mathcal{K}'}) ( \mathbb{F}^{AA'} \otimes \mathbb{I}^{\bar{A}\bar{A}'})].
\end{equation}

It is straightforward that
\begin{align}
\tr[ \mathbb{I}^{\mathcal{K}\mathcal{K}'} ( \mathbb{F}^{AA'} \otimes \mathbb{I}^{\bar{A}\bar{A}'})]
&=
\tr[ \Pi_{\mathcal{K}}^{\Lambda} \otimes \Pi^{\Lambda}_{\mathcal{K}'}( \mathbb{F}^{AA'} \otimes \mathbb{I}^{\bar{A}\bar{A}'})]\\
&=
D_{\mathcal{K}}^2 \tr[ (\Omega_{\mathcal{K}}^A \otimes \Omega_{\mathcal{K}'}^A ) \mathbb{F}^{AA'}]\\
&=
D_{\mathcal{K}}^2 \tr[ (\Omega_{\mathcal{K}}^A)^2].
\end{align}
To compute $\tr[ \mathbb{F}^{\mathcal{K}\mathcal{K}'} ( \mathbb{F}^{AA'} \otimes \mathbb{I}^{\bar{A}\bar{A}'})]$, we expand the swap operator $\mathbb{F}^{\mathcal{K}\mathcal{K}'}$ as
\begin{equation}
\mathbb{F}^{\mathcal{K}\mathcal{K}'} = \sum_{i, j =1}^{D_{\mathcal{K}}} \ketbra{\psi_i}{\psi_j}^{\Lambda} \otimes \ketbra{\psi_j}{\psi_i}^{\Lambda'},
\end{equation}
where $\{ \ket{\psi_i} \}_i$ is an orthonormal basis in $\mathcal{K}$. This allows us to explicitly write down \mbox{$\tr[ \mathbb{F}^{\mathcal{K}\mathcal{K}'} ( \mathbb{F}^{AA'} \otimes \mathbb{I}^{\bar{A}\bar{A}'})]$ as}
\begin{equation}
\tr[ \mathbb{F}^{\mathcal{K}\mathcal{K}'} ( \mathbb{F}^{AA'} \otimes \mathbb{I}^{\bar{A}\bar{A}'})]
=
\sum_{i, j =1}^{D_{\mathcal{K}}} \tr\bigl[ \tr_{\bar{A}} [\ketbra{\psi_i}{\psi_j}^{\Lambda} ] \tr_{\bar{A}} [\ketbra{\psi_j}{\psi_i}^{\Lambda} ] \bigr],
\end{equation}
where we used the swap trick.
We further expand $\ket{\psi_i}$ as $\sum_{\alpha =1}^{d^{n_{\bar{A}}}} \ket{\tilde{\psi}_{i}^{\alpha}}^A \otimes \ket{\alpha}^{\bar{A}}$ by using an orthogonal basis $\{ \ket{\alpha} \}_{\alpha}$ in $\mathcal{H}^{\bar{A}}$.
Note that $\ket{\tilde{\psi}_{i}^{\alpha}}^A := (I^A \otimes \bra{\alpha}^{\bar{A}}) \ket{\psi_i}$ are un-normalized. Using this notation, we~have
\begin{equation}
\tr_{\bar{A}} [\ketbra{\psi_i}{\psi_j}^{\Lambda} ] 
=
\sum_{\alpha =1}^{d^{n_{\bar{A}}}} \ket{\tilde{\psi}_{i}^{\alpha}}\! \langle \tilde{\psi}_{j}^{\alpha}|^A,
\end{equation}
leading to
\begin{equation}
\tr[ \mathbb{F}^{\mathcal{K}\mathcal{K}'} ( \mathbb{F}^{AA'} \otimes \mathbb{I}^{\bar{A}\bar{A}'})]
=
\sum_{i, j =1}^{D_{\mathcal{K}}} 
\sum_{\alpha, \beta =1}^{d^{n_{\bar{A}}}} \langle \tilde{\psi}_{i}^{\beta} \ket{\tilde{\psi}_{i}^{\alpha}}  \langle \tilde{\psi}_{j}^{\alpha}\ket{\tilde{\psi}_{j}^{\beta}}.
\end{equation}
We then use the relation that $\langle \tilde{\psi}_{i}^{\beta} \ket{\tilde{\psi}_{i}^{\alpha}} =\bra{\psi_i} (I^A \otimes\ket{\beta} \! \langle \alpha|^{\bar{A}}) \ket{\psi_i} = \langle \alpha| \psi_i^{\bar{A}} \ket{\beta}$ and obtain
\begin{align}
\tr[ \mathbb{F}^{\mathcal{K}\mathcal{K}'} ( \mathbb{F}^{AA'} \otimes \mathbb{I}^{\bar{A}\bar{A}'})]
&=
\sum_{\alpha, \beta =1}^{d^{n_{\bar{A}}}} \biggl| \sum_{i =1}^{D_{\mathcal{K}}}  \langle \alpha| \psi_i^{\bar{A}} \ket{\beta} \biggr|^2\\
&=
\sum_{\alpha, \beta =1}^{d^{n_{\bar{A}}}} \biggl| \tr \biggl( I^A \otimes \langle \alpha|^{\bar{A}}\biggr) \biggl( \sum_{i =1}^{D_{\mathcal{K}}}   \psi_i^{\Lambda} \biggr) \biggl( I^A \otimes \ket{\beta}^{\bar{A}} \biggr) \biggr|^2\\
&=
\sum_{\alpha, \beta =1}^{d^{n_{\bar{A}}}} \biggl| \tr \bigl( I^A \otimes \langle \alpha|^{\bar{A}}\bigr) \Pi^{\mathcal{K}} \bigl( I^A \otimes \ket{\beta}^{\bar{A}} \bigr) \biggr|^2\\
&=
D_{\mathcal{K}}^2 \sum_{\alpha, \beta =1}^{d^{n_{\bar{A}}}} \bigl| \langle \alpha|\Omega_{\mathcal{K}}^{\bar{A}} \ket{\beta} \bigr) \bigr|^2\\
&=
D_{\mathcal{K}}^2 \tr (\Omega_{\mathcal{K}}^{\bar{A}})^2 
\end{align}

Altogether, we have
\begin{equation}
\mathbb{E}E_A(\ket{\phi}) \geq 
-\log[ \mathbb{E} \tr[(\phi^A)^2] ]
= -\log \biggl[ \biggl( 1- \frac{1}{D_{\mathcal{K}}+1} \biggr)\bigl(  \tr  (\Omega_{\mathcal{K}}^{A})^2  +  \tr (\Omega_{\mathcal{K}}^{\bar{A}})^2 \bigr) \biggr] = \bar{S}(\mathcal{K}).
\end{equation}
Substituting this into Equation~\eqref{Ineq:Levy}, we obtain the desired statement. $\hfill$
\end{proof}

Theorem~\ref{Thm:ConEnt} implies that when $D_{\mathcal{K}} \gg (\ln R_{\mathcal{K}})^2$, the~entanglement entropy of a random state $\ket{\phi} \sim {\sf H}_{\mathcal{K}}$ in the subspace $\mathcal{K}$ is typically more than $\bar{S}(\mathcal{K})$. Hence, when we are interested in the entanglement entropy of a random state in the subspace $\cK$, what we need to do is to compute $\bar{S}(\mathcal{K})$, $D_{\mathcal{K}}$, and~$R_{\mathcal{K}}$. 

\section{Generic Entanglement of States with an Axial~Symmetry} \label{S:AxialSymmetry}

Based on Theorem~\ref{Thm:ConEnt}, we now study generic entanglement when a random state has symmetry. 
We start with a simple case of an axial symmetry of qubit-systems. {This is because the relation between generic entanglement and quantum error correcting codes~\cite{NWK2020} is particularly pointed out when the system has an axial symmetry.}

Suppose that the system consists of $n$-qubits and has an axial symmetry. Without~loss of generality, we assume that the symmetry is around the $Z$-axis. Each invariant subspace is then characterized by the $Z$-component of angular momentum, or~equivalently, the~number $m$ of up-spins as follows:
\begin{equation}
\mathcal{H}^{\Lambda} = \bigoplus_{m=0}^{n} \mathcal{H}^{\Lambda}_m,
\end{equation}
where $\mathcal{H}^{\Lambda}_m = {\rm span} \bigl\{ \ket{\phi}: S_Z \ket{\phi} = (m - n/2) \ket{\phi} \bigr\}$ with $S_Z$ being the spin-$Z$ operator on $n$ qubits, i.e.,~$S_Z = \sum_{i=1}^n S_Z^{(i)}$ with $S_Z^{(i)}$ being the spin-$Z$ operators acting on the $i$th qubit.
The dimension $D_m$ of each subspace $\mathcal{H}^{\Lambda}_m$ is given by $D_m = \binom{n}{m}$.
We consider the entanglement entropy $E_A(\ket{\phi})$ of a random state $\ket{\phi} \sim {\sf H}_m$, where ${\sf H}_m$ is the Haar measured on the subspace $\cH_m^{\Lambda}$.

Since each subspace $\mathcal{H}^{\Lambda}_m$ can be spanned by the basis consisting of product states, $\Omega_{\mathcal{K}}^A$ for $\mathcal{K} = \mathcal{H}_m^{\Lambda}$, which we simply denote by $\Omega_{m}^A$, can be simply obtained as
\begin{equation}
\Omega_{m}^A  = \frac{1}{\binom{n}{m}} \sum_{\ell = 0}^m  \binom{n_{\bar{A}}}{m-\ell} \Pi^{A}_{\ell},
\end{equation}
where $\Pi^{A}_{\ell}$ is the projection onto the subspace of $\mathcal{H}^A$ spanned by the states with $\ell$ up-spins. 
\mbox{We similarly have}
\begin{equation}
\Omega_{m}^{\bar{A}}  = \frac{1}{\binom{n}{m}} \sum_{\ell = 0}^m  \binom{n_A}{m-\ell} \Pi^{\bar{A}}_{\ell}.
\end{equation}
Thus, the~$\bar{S}(\mathcal{H}_m)$ is given by
\begin{equation}
\bar{S}(\mathcal{H}_m)
=
-\log \biggl[ \sum_{\ell = 0}^m   \biggl( \frac{\binom{n_{\bar{A}}}{m-\ell}}{\binom{n}{m}}\biggr)^2 \binom{n_A}{\ell} + \sum_{\ell = 0}^m   \biggl( \frac{\binom{n_A}{m-\ell}}{\binom{n}{m}}\biggr)^2 \binom{n_{\bar{A}}}{\ell} \biggr]
-
\log \bigl[ 1 - \frac{1}{\binom{n}{m} +1} \bigr].
\end{equation}

\begin{figure}[h]
\begin{center}
\includegraphics[width=135mm]{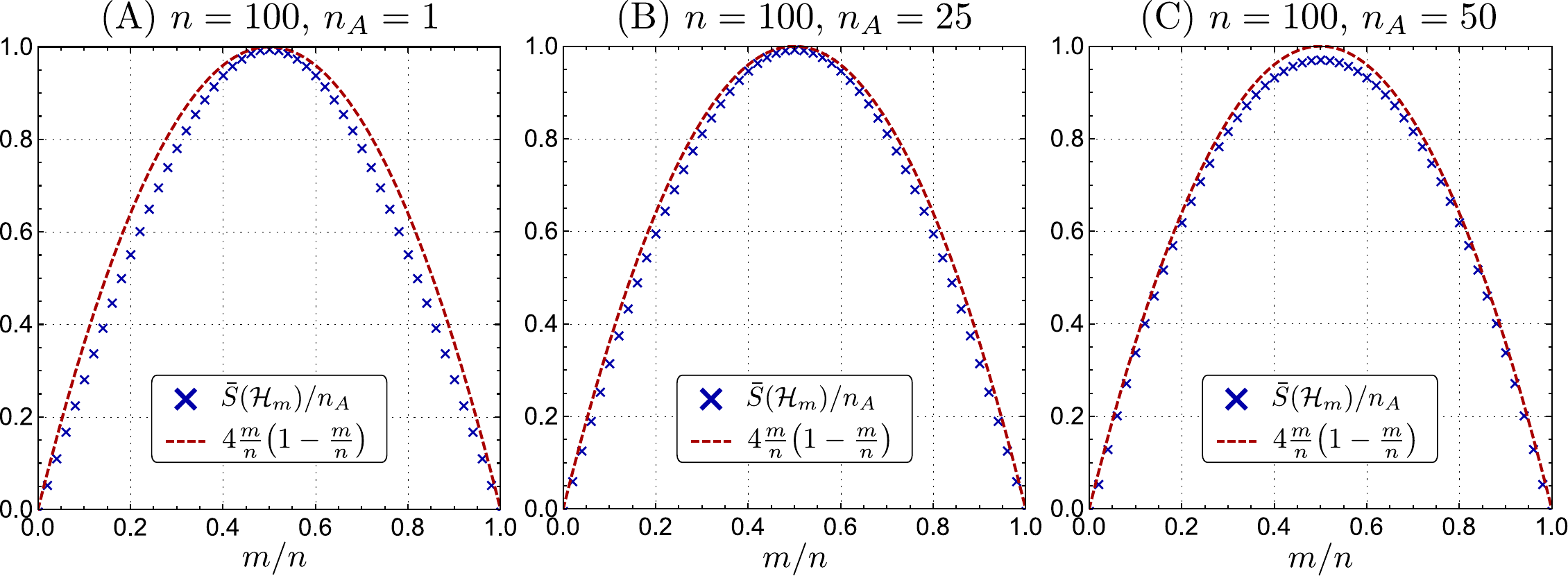}
\caption{The function $\bar{S}(\cH_m)/n_A$ is plotted by $\times$ as a function of $m/n$ for a $n=100$ and $n_A=1$ (\textbf{A}), $n_A = 25$ (\textbf{B}), and~$n_A = 50$ (\textbf{C}). We also provide a function $f(m/n):=4m/n(1- m/n)$ by a red dashed line in each figure. It is clear that $\bar{S}(\cH_m)/n_A \approx f(m/n)$ for any $n_A$ and $m$.}
\label{Fig:ZSym}
\end{center}
\end{figure}

In Figure~\ref{Fig:ZSym}, we plot $\bar{S}(\cH_m)/n_A$ as a function of $m/n$ for a fixed $n$, and~also the function \mbox{$f(m/n):= 4m/n(1- m/n)$}. Since they coincide well, we approximate $\bar{S}(\cH_m)$ by a quadratic function.
\begin{equation}
\bar{S}(\cH_m) \approx  4\frac{m}{n} \bigl(1- \frac{m}{n} \bigr) n_A.
\end{equation}

Using this expression and denoting $m$ by $\gamma n$ with $\gamma \in [0, 1]$, we obtain from Theorem~\ref{Thm:ConEnt} that a random state $\ket{\phi} \sim {\sf H}_{\gamma n}$ in the subspace $\cH_{\gamma n}^{\Lambda}$ with a fixed $Z$-axis angular momentum \mbox{$(\gamma - 1/2) n$ satisfies}
\begin{equation}
{\rm Prob}_{\ket{\phi} \sim {\sf H}_m} \bigl[ E_A(\ket{\phi})  \geq 4\gamma(1 - \gamma) n_A  - \epsilon \bigl]
>
1-  \exp \biggl[- \frac{(\binom{n}{m} +1) \epsilon^2}{72 (\pi \ln 2)^3 n_A^2} \biggr], \label{Eq:38}
\end{equation}
for any $\epsilon>0$. Note that we used a trivial bound $2^{n_A}$ on $R_{\mathcal{K}} = \max_{\ket{\phi} \in \mathcal{K}}[ {\rm supp} (\phi^A) ]$.
This implies that as far as $\gamma$ is constant, the~state still obeys the volume law, i.e.,~the entanglement entropy is proportional to the number of qubits $n_A$ in the subsystem $A$. In~this sense, the~axial symmetry does not change the volume law of entanglement. However, recalling that the entanglement entropy of a Haar random state of qubits without any symmetry is $\approx$$n_A$, the~axial symmetry can reduce the entanglement entropy by a constant degree since Equation~\eqref{Eq:38} shows that the coefficient of $n_A$ is $4 \gamma (1-\gamma)$ that can be smaller than $1$.

\unskip
\vspace{-6pt}

\section{Generic Entanglement of States with Permutation~Symmetry}  \label{S:PermSym}
We next investigate the entanglement entropy of a random state with the permutation symmetry.
{It is often argued that a state with permutation symmetry is generally weakly entangled~\cite{HMMOV2008, CSW2012}. Based on Theorem~\ref{Thm:ConEnt}, we here quantitatively justify that this common belief indeed holds for most permutation symmetric states.}

 We especially consider the symmetric and antisymmetric subspaces in $\mathcal{H}^{\Lambda}$ of $n$ qudits, which are respectively defined by
\begin{align}
&\mathcal{H}_{+}^{\Lambda} := {\rm span} \bigl\{ \ket{\phi} \in \mathcal{H}^{\Lambda} : U_{\sigma}\ket{\phi} =  \ket{\phi}, \forall \sigma \in \mathcal{P}_n   \bigr\}, \\
&\mathcal{H}_{-}^{\Lambda} := {\rm span} \bigl\{ \ket{\phi} \in \mathcal{H}^{\Lambda} : U_{\sigma} \ket{\phi} = {\rm sign}(\sigma)  \ket{\phi}, \forall \sigma \in \mathcal{P}_n \bigr\},
\end{align}
where $\mathcal{P}_n$ is the permutation group of degree $n$, and~$U_{\sigma}$ is a unitary representation of $\sigma \in \mathcal{P}_n$. 
The~dimensions $D_{\pm}$ of $\mathcal{H}_{\pm}^{\Lambda}$ are given by
$D_+ = \binom{n+d-1}{n}$ and $D_- = \binom{d}{n}$, respectively.
Note that $\mathcal{H}_{-}^{\Lambda}$ becomes non-trivial if and only if $n \leq d$. 
From the physics point of view, the~symmetric (antisymmetric) subspace is a Hilbert space of indistinguishable bosons (fermions). 

Let us first consider the entropy of a state $\Omega^A_{\pm} =  \tr_{\bar{A}} [\Pi_{\pm}^{\Lambda}/D_{\pm} ]$, where $\Pi_{\pm}^{\Lambda}$ is the projection onto the symmetric/anti-symmetric subspace in $\mathcal{H}^{\Lambda}$.
Due to the special properties of the permutation symmetry, it turns out that $\Omega^A_{\pm} =  \Pi_{\pm}^A/\tr[\Pi_{\pm}^A]$.
To see this, we use another expression of $\Pi_{\pm}^{\Lambda}/D_{\pm}$, that~is
\begin{equation}
\frac{\Pi_{\pm}^{\Lambda}}{D_{\pm}} = \int_{\mathcal{U}(d)} u^{\otimes n} \ketbra{\varphi_{\pm}}{\varphi_{\pm}} u^{\dagger \otimes n}  du,
\end{equation}
where $u \in \mathcal{U}(d)$ in the integral is a unitary acting on a single qudit, $\cU(d)$ is the unitary group of degree $d$, and~$\ket{\varphi_{\pm}}$ is any state in $\mathcal{H}_{\pm}^{\Lambda}$. This is a consequence of Schur's lemma~\cite{SW1986} and the fact that the symmetric and anti-symmetric subspaces are irreducible representations of $\mathcal{U}(d)$ that acts as $u^{\otimes n}$ \mbox{onto $\mathcal{H}^{\Lambda}$}.

For the symmetric subspace, we can take $\ket{\varphi_+}$ as a product state $\ket{0}^{\otimes n}$. Then, we have
\begin{align}
\Omega^A_{+} &=  \tr_{\bar{A}} \bigl[\Pi_{+}^{\Lambda}/D_{\pm} \bigr]\\
&=\tr_{\bar{A}} \biggl[\int_{\mathcal{U}(d)} u^{\otimes n} \ketbra{0}{0}^{\otimes n} u^{\dagger \otimes n}  du \biggr]\\
&=\int_{\mathcal{U}(d)} \tr_{\bar{A}} \bigl[u^{\otimes n} \ketbra{0}{0}^{\otimes n} u^{\dagger \otimes n} \bigr]  du\\
&=\int_{\mathcal{U}(d)} u^{\otimes n_A} \ketbra{0}{0}^{\otimes n_A} u^{\dagger \otimes n_A}  du\\
&=\frac{\Pi_{+}^{A}}{\tr[\Pi_{+}^{A}]}.
\end{align}
In the last line, we again used Shur's lemma and the fact that $\ketbra{0}{0}^{\otimes n_A} \in \mathcal{H}^A_+$.
For the anti-symmetric subspace, we similarly obtain
\begin{align}
\Omega^A_- &= \int_{\mathcal{U}(d)} \tr_{\bar{A}} \bigl[u^{\otimes n} \ketbra{\varphi_-}{\varphi_-} u^{\dagger \otimes n} \bigr]  du \\
&= \int_{\mathcal{U}(d)}  u^{\otimes n_A} \tr_{\bar{A}} \bigl[\ketbra{\varphi_-}{\varphi_-} \bigr] u^{\dagger \otimes n_A}  du.
\end{align}
To check the support of $\tr_{\bar{A}} [\ketbra{\varphi_-}{\varphi_-} ]$, we decompose $\ket{\varphi_-}$ into the form of $\sum_{i} \ket{\phi_i}^{A} \otimes \ket{i}^{\bar{A}}$, where $\ket{i}^{\bar{A}} = \ket{i_1}\otimes \dots \otimes \ket{i_{n_{\bar{A}}}}$ ($i_{\ell} = 0, \dots, d-1$) is an orthonormal product basis in $\bar{A}$. For~any $i$, the~state $\ket{\phi_i}^{A}$ should be also anti-symmetric because, for~any permutation $\sigma \in \mathcal{P}_{n_A}$, \mbox{$(U_{\sigma}^A \otimes I^{\bar{A}}) \ket{\varphi_-} = {\rm sign} (\sigma) \ket{\varphi_-} = {\rm sign} (\sigma)\sum_{i} \ket{\phi_i}^{A} \otimes \ket{i}^{\bar{A}}$}. Recalling that $\ket{\phi_i}^{A} = (I^A \otimes \bra{i}^{\bar{A}}) \ket{\varphi_-}$, we~obtain $U_{\sigma}^A \ket{\phi_i}^A = {\rm sign} (\sigma) \ket{\phi_i}$, implying that $\ket{\phi_i} \in \mathcal{H}^A_-$ for any $i$. 
Thus, the~support of $\tr_{\bar{A}}[ \ketbra{\phi_-}{\phi_-}]$ is $\cH_-^{\bar{A}}$. Again using the Schur's lemma and the fact that the anti-symmetric subspace is irreducible, we~obtain
\begin{equation}
\Omega^A_-=\frac{\Pi_-^{A}}{\tr[\Pi_-^{A}]}.
\end{equation}

It is now straightforward to compute $\tr (\Omega^X_{\pm})^2$ for $X=A, \bar{A}$ as $\tr (\Omega^X_{\pm})^2 = 1/D_{\pm}^X$, where $D_{\pm}^X = \tr[\Pi_{\pm}^X]$, leading to
\begin{align}
\bar{S}(\mathcal{H}_{\pm}) &= -\log \biggl[ \frac{1}{D_{\pm}^A} + \frac{1}{D_{\pm}^{\bar{A}}}  \biggr]
-\log \biggl[ 1 - \frac{1}{D_{\pm}+1} \biggr]\\
&\approx
\log[D_{\pm}^A] -  \frac{D_{\pm}^A}{D_{\pm}^A + D_{\pm}^{\bar{A}}}.
\end{align}
We also have $R_{\mathcal{H}_{\pm}} = D^A_{\pm}$ since, by~taking the partial trace, symmetric and anti-symmetric states remain in the symmetric and anti-symmetric subspaces, respectively. Note that this also guarantees that trivial upper bounds of the entanglement entropy for symmetric/anti-symmetric states are given by $\log[D_{\pm}^A]$.

From Theorem~\ref{Thm:ConEnt}, we finally obtain the following: for any $\epsilon >0$, a~random state $\ket{\phi} \sim {\sf H}_{\pm}$ in the symmetric/anti-symmetric subspace in $\mathcal{H}^{\Lambda}$ satisfies
\begin{equation}
{\rm Prob}_{\ket{\phi} \sim {\sf H}_{\pm}} \biggl[\log D_{\pm}^A \geq  E_A(\ket{\phi}) \geq \log D_{\pm}^A -  \frac{D_{\pm}^A}{D_{\pm}^A + D_{\pm}^{\bar{A}}} -\epsilon
\biggr]
>
1-  \exp \biggl[- \frac{(D_{\pm} +1) \epsilon^2}{72 \pi^3 \ln 2 (\ln D_{\pm}^A)^2} \biggr].
\end{equation}
Since $D_{+} = \binom{n+d-1}{d-1} \gg D^A_{+} = \binom{n_A+d-1}{d-1}$ and $D_{-} = \binom{d}{n} \gg D^A_{-} = \binom{d}{n_A}$ when $n_A \ll n$, the~right-hand side is extremely close to $1$. Hence, we conclude that the entanglement entropy for symmetric/anti-symmetric random states in a small subsystem $A$ extremely concentrates between $\log D_{\pm}^A - D_{\pm}^A/(D_{\pm}^A + D_{\pm}^{\bar{A}})$ and $\log D_{\pm}^A$.

To be more concrete, let us consider special cases of $d$. For~simplicity, we ignore $D^A_{\pm}/(D_{\pm}^A + D_{\pm}^{\bar{A}})$. We first look at the entanglement entropy of a random state in the symmetric space, which typically takes the following value:
\begin{equation}
E_A(\ket{\phi}) \approx
\begin{cases}
(d-1) \ln [n_A + 1] & \text{when\ } d \ll n_A,\\
(n_A + d-1) H\bigl( \frac{d-1}{n_A + d-1} \bigr) & \text{when\ } d = \Theta(n_A),\\
n_A \ln[d-1] & \text{when\ } n_A \ll d,
\end{cases}
\end{equation}
where $H(p) = -p \log p - (1-p) \log (1-p)$ for $0 \leq p \leq 1$ is the binary entropy. 
We especially note that when $d \ll n_A$, the~entanglement entropy $E_A(\ket{\phi})$ for a random symmetric state is typically~$\approx$$\log n_A$, and~fails to satisfy the volume law of entanglement. Thus, our result implies that the volume law fails to hold when the many-body system is permutation symmetric and consists of the particles with a constant degree of freedom, so that $d \ll n_A$. A~simple example may be many-body systems composed of indistinguishable bosons.
We however note that this result is a consequence of the facts that symmetric states remain symmetric by taking the partial trace and that the symmetric subspace is~small.

On the other hand, for~the anti-symmetric random states, we have
\begin{equation}
E_A(\ket{\phi}) \approx d H(\gamma),
\end{equation}
where $\gamma:=n_A/d$. Note that $\gamma \leq 1$ since the anti-symmetric space is non-trivial only when $n \leq d$. Since~the entanglement entropy of the random state without any symmetry is typically $\gamma d \log d$ in terms of $\gamma$, we conclude that the anti-symmetric condition of the state typically reduces the entanglement entropy by the factor $\Theta(\log d)$.

\section{Generic Entanglement of States with Translation~Symmetry} \label{S:TransSym}

As the last, but~not least, instance of symmetry, we consider translation symmetry, which~is one of the most common symmetries in many-body systems.
We especially consider the case where qudits are aligned on a one-dimensional line with the periodic boundary condition and $A$ is an interval of the line. The~corresponding group $\mathcal{T}$ is generated by the shifting operator $T$, which~shifts every qudit to the next site.
Since $T^n$ is the identity due to the periodic boundary condition, the~Hilbert space $\mathcal{H}^{\Lambda}$ is decomposed into discrete momentum subspaces such as \mbox{$\mathcal{H}^{\Lambda} =\bigoplus_{\theta} \mathcal{H}_{\theta}^{\Lambda}$}, where~\mbox{$\theta \in \{ \frac{2 \pi k}{n} : k = 0, \dots, n-1 \}$}.
Here, each subspace is defined by
\begin{equation}
\mathcal{H}_{\theta}^{\Lambda} := {\rm span} \bigl\{ \ket{\phi} \in \mathcal{H}^{\Lambda} : U_T \ket{\phi} = e^{i \theta}  \ket{\phi}   \bigr\},
\end{equation}
and $U_T$ is a unitary representation of $T$. This decomposition corresponds to a discrete version of Bloch's theorem.
For simplicity, we consider only the case where $n$ is a prime number. This simplifies the analysis, but~we expect that nearly the same result holds even when $n$ is not prime with a slight~modification.
 
To investigate the entanglement entropy, we first provide a basis in $\mathcal{H}_{\theta}^{\Lambda}$ and explicitly write down the projector $\Pi_{\theta}^{\Lambda}$ onto the subspace.
Let $\hat{\mathcal{C}}$ be the set of $n$-dit sequences, \mbox{$\hat{\mathcal{C}}=\{ 0\dots00, 0\dots01,\dots,d-1\dots d-1\}$}, and~$\mathcal{C}$ be the set $\hat{\mathcal{C}} \setminus \{ \vec{a}\}_{a}$, where~$\vec{a}$ for $a = 0, \dots, d-1$ is the $n$-dit sequence whose components are all $a$.
Let $\mathcal{C}_{\mathcal{T}}$ be an equivalent class of $\mathcal{C}$ by the translation group $\mathcal{T}$, $\mathcal{C}_{\mathcal{T}}:=\mathcal{C} / \mathcal{T}$.
We construct a basis in $\mathcal{H}_{\theta}^{\Lambda}$ using the state
\begin{equation}
\ket{c_\theta} = \frac{1}{\sqrt{n}} \sum_{k=0}^{n-1} e^{i \theta k }  U_T^{k} \ket{c},
\end{equation}
for $c \in \mathcal{C}_{\mathcal{T}}$. Note that none of $\ket{c_\theta}$ is a zero vector due to the assumption that $n$ is a prime number.
The~basis is given by
\begin{equation}
\mathcal{B}_{0}^{\Lambda} = 
\{ \ket{\vec{a}} \}_{a=0, 1,\dots,d-1} \cup \{ \ket{c_0} \}_{c \in \mathcal{C}_{\mathcal{T}}},
\end{equation}
and, for~$\theta \neq 0$,
\begin{equation}
\mathcal{B}_{\theta}^{\Lambda} = 
 \{ \ket{c_\theta} \}_{c \in \mathcal{C}_{\mathcal{T}}}.
\end{equation}
Clearly, the~dimension $D_{\theta}^{\Lambda} $ of each subspace is given by
\begin{equation}
D_{\theta}^{\Lambda} = 
\begin{cases} 
\frac{d^n -d}{n} +d & \text{ for } \theta = 0, \\  
\frac{d^n -d}{n} & \text{ otherwise}.
\end{cases}
\end{equation}

Using these bases, we derive a upper bound of $\tr  (\Omega_{\theta}^X)^2$ for $X=A, \bar{A}$, from~which we obtain a lower bound of $-\log[ \tr  (\Omega_{\theta}^A)^2 + \tr  (\Omega_{\theta}^A)^2]$. Since $A$ and $\bar{A}$ can be treated in the same way, we consider only $\Omega_{\theta}^A$, which can be expanded as
$\Omega_{\theta}^A
=
\sum_{a,b}
\omega^{\theta}_{ab }
\ketbra{a}{b},$
where $a =a_1 \dots a_{n_A}$ and $b=b_1\dots b_{n_A}$ ($a_i,b_i \in \{0,1,\dots, d-1\}$ for all $i=1,\dots, n_A$). 
The off-diagonal terms $\omega_{a b}^{\theta}$ ($a \neq b$) are non-zero if and only if there exists $v =v_1 \dots v_{n_{\bar{A}}} $ ($v_i \in \{ 0, \dots, d-1\}$ for $i=1,\dots,n_{\bar{A}}$) such that
\begin{equation}
\ket{a \oplus v}  = U_T^k \ket{b \oplus v} \label{Eq:a}
\end{equation}
for some $k \in \{1,\dots, n-1\}$. Here, we used the notation that $a \oplus v = a_1 \dots a_{n_A} v_1 \dots v_{n_{\bar A}}$.
Hence, if~the number of $i$'s ($i=0,\dots, d-1$) in $a$ differs from that in $b$, $\omega_{ab}^{\theta} =0$.
This means that $\Omega_{\theta}^A$ is decomposed into positive operators $\Omega_{\theta}^A (m_0, \dots, m_{d-1})$ on the Hilbert spaces spanned by states with configurations $c$ containing $m_i$ of $i$'s ($i=0,\dots, d-1$);
\begin{equation}
\Omega_{\theta}^A = \bigoplus_{(m_0, \dots, m_{d-1})} \Omega_{\theta}^A (m_0, \dots, m_{d-1}),
\end{equation}
where $m_i$ runs from $0$ to $n_A$ under the condition that $\sum_{i=0}^{d-1} m_i = n_A$.
Thus $\tr (\Omega_{\theta}^A)^2$ is given by
\begin{equation}
\tr (\Omega_{\theta}^A)^2 = \sum_{(m_0, \dots, m_{d-1})} \tr \bigl[ \bigl( \Omega_{\theta}(m_0, \dots, m_{d-1}) \bigr)^2 \bigr]. \label{Eq:Decomp}
\end{equation}
The dimension of the support of $\Omega_{\theta}^A (m_0, \dots, m_{d-1})$ is $M(m_0, \dots, m_{d-1}) := \frac{n_A!}{m_0! \dots m_{d-1}!}$.

From a counting argument, the~diagonal terms in $\Omega_{\theta}^A$ are obtained as
\begin{equation}
\omega_{aa}^{\theta}
=\begin{cases}
\frac{d^{n_{\bar{A}}} + m_{\theta}}{n D_{\theta}^{\Lambda}} & \text{ for } a = \vec{0},\vec{1},\dots,\overrightarrow{d-1},\\
\frac{d^{n_{\bar{A}}}}{n D_{\theta}^{\Lambda}} & \text{ otherwise },
\end{cases} \label{Eq:Diag}
\end{equation}
where $m_{\theta}= n \delta_{\theta 0}-1$ with $\delta_{\theta 0}$ being the delta function. 
For the off-diagonal terms, we show that the absolute value of each of them is not greater than $1/D_{\theta}^{\Lambda}$.
For a fixed $a$, $b$, and~$k$, there exists at most one $v$ that satisfies Equation~\eqref{Eq:a} due to the assumption that $n$ is a prime number.
Recalling that $k \neq 0$ since $a \neq b$, an~off-diagonal term of $\Omega_{\theta}(m_0, \dots, m_{d-1})$ is a summation of at most $n-1$ terms, where each term has coefficient $e^{i \theta p}/(n D_{\theta}^{\Lambda})$ for some $p \in \{1,\dots, n-1 \}$.
Thus, all off-diagonal terms of $\Omega_{\theta}(m_0, \dots, m_{d-1})$ are bounded from above by
\begin{equation}
\frac{1}{n D_{\theta}^{\Lambda}} \biggl| \sum_{x =1}^{n-1} q_x e^{i \theta p_x} \biggr| \leq \frac{1}{D_{\theta}^{\Lambda}}, \label{Eq:Offdiag}
\end{equation}
where $q_x \in \{0,1\}$ is an indicator function that $q_x =1$ if there exists $v$ satisfying Equation~\eqref{Eq:a} for $k=x$ and $q_x=0$ otherwise.

By substituting the diagonal terms, Equation~\eqref{Eq:Diag}, and~the upper bounds of off-diagonal terms, Equation~\eqref{Eq:Offdiag}, into~Equation~\eqref{Eq:Decomp}, we obtain an upper bound of $\tr (\Omega_{\theta}^A)^2$ as
\begin{equation}
\tr (\Omega_{\theta}^A )^2 \leq 
\frac{1}{d^{n_A} (1+m_{\theta} d^{1-n})^2}
\biggl( 1+\frac{2 m_{\theta}}{d^{n-1}}+\frac{m_{\theta}^2 d + n^2 \Gamma^A}{d^{n+n_{\bar{A}}}} \biggr), \label{Eq:UpperB}
\end{equation}
where
\begin{equation}
\Gamma_A= \sum_{(m_0,\dots, m_{d-1})} \biggl[ \biggl( \frac{n_A!}{m_0! \dots m_{d-1}!} \biggr)^2 - \frac{n_A!}{m_0! \dots m_{d-1}!} \biggr].
\end{equation}
As $d^{n_A} < \Gamma_A< d^{2n_A}$,  we have
\begin{equation}
\tr [ (\Omega_{\theta}^A )^2 ] \leq 
d^{-n_A} ( 1+n^2 d^{-n_{\bar{A}}} ) + o(d^{-n}).
\end{equation}
Similarly, we can derive an upper bound for $\tr [ (\Omega_{\theta}^{\bar{A}} )^2]$ as
\begin{align}
\tr [ (\Omega_{\theta}^{\bar{A}} )^2] 
&\leq 
d^{-n_{\bar{A}}} ( 1+n^2 d^{-n_A} ) + o(d^{-n}).
\end{align}

Based on these lower bounds, we obtain
\begin{equation}
\bar{S}(\mathcal{H}_{\theta}^{\Lambda}) \geq  n_A \log d - d^{-n+2 n_A} + o(d^{-n+n_A}).
\end{equation}
Using a trivial upper bound $d^{n_A}$ on $R_{\mathcal{H}_{\theta}^{\Lambda}}$ for any $\theta$, we arrive at our conclusion: for any $\epsilon >0$ and for $n/2 \geq n_A$, it holds that
\begin{equation}
{\rm Prob}_{\ket{\phi} \sim {\sf H}_{\theta}} \bigl[ E_A(\ket{\phi})  \geq n_A \log d - d^{-n + 2 n_A} + o(d^{-n+n_A})- \epsilon \bigl]
>
1-  \exp \biggl[- \frac{C \epsilon^2}{72 \pi^3 \ln 2} \biggr],
\end{equation}
for any $\theta$, where $C = O(d^n/(n n_A^2))$. Since the entanglement entropy for any state is bounded from above by $n_A \log d$, this implies that the entanglement entropy of a random state with translation symmetry concentrates between $n_A \log d - d^{-n + 2 n_A}$ and $n_A \log d$.
Hence, translation symmetry changes the generic entanglement only~slightly.

\section{Entanglement Phases and~Symmetries} \label{S:EntPhase}

We finally investigate how symmetries affect the entanglement phases. The~original analyses of entanglement phases are based on the technique of the random matrix theory~\cite{G2007,FMPPS2008,PFPPS2010,NMV2010,PFGPPS2011,NMV2011,FFPPY2013,FPPSY2019}. We here present numerical calculations of the entanglement entropy of random states in various invariant subspaces associated with symmetry. We especially consider a random symmetric state, and~a random translation invariant state for $d=2$, $n=10$, and~$n_A = 5$. All numerics are done by sampling pure states from an invariant subspace of the symmetry. We have used the so-called Hurwitz parametrization of a state, based on which a parametrization of a Haar random state is known~\cite{PZK1998}.

\begin{figure}[tb!]
\begin{center}
\includegraphics[width=155mm]{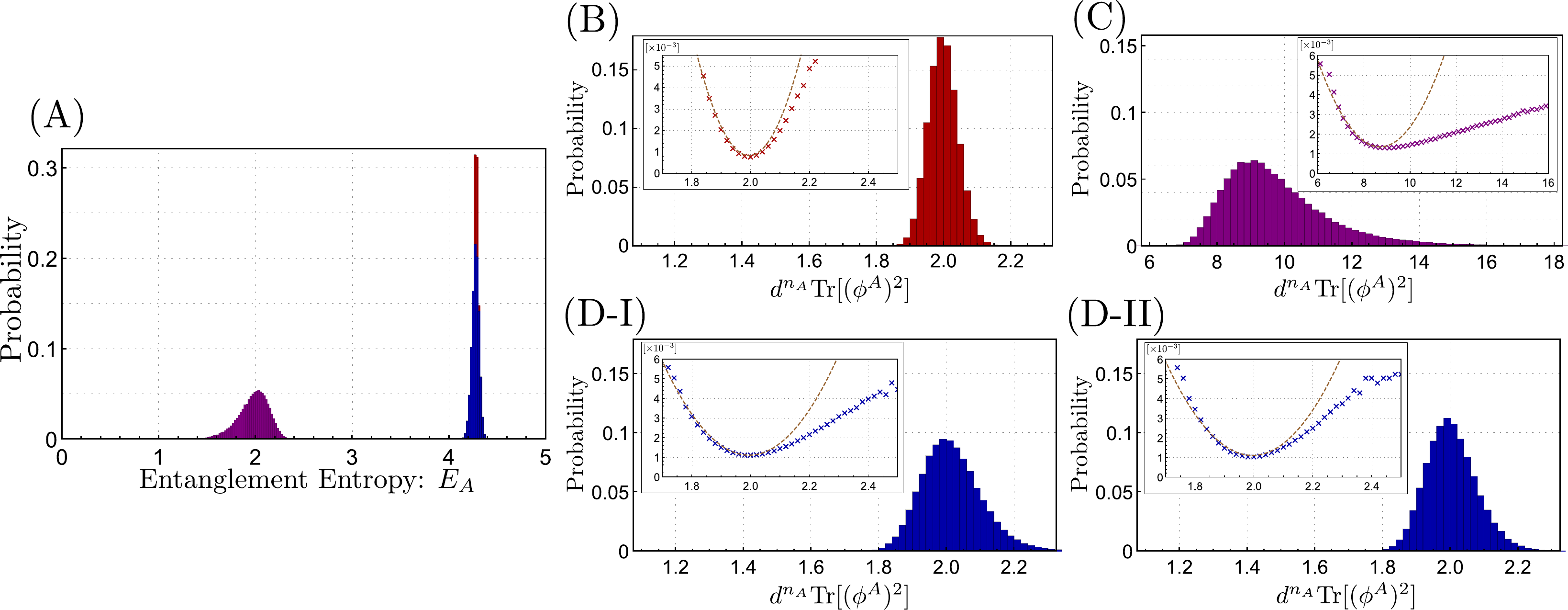}
\caption{The distributions of entanglement over the random states without/with symmetry, which~are numerically obtained for $d=2$, $n=10$, and~$n_A=5$. The~number of samples is $10^5$, binned in intervals of $0.02$ for Panels (\textbf{A},\textbf{B},\textbf{D-I}, \textbf{D-II}), $0.2$ for Panel (\textbf{C}).
Panel (\textbf{A}) shows the distribution of the entanglement entropy $E_A$ over a Haar random state without symmetry (red), that over a random symmetric state (purple), and~that over a random translation invariant state for $\theta = 0$ (blue). We observe that only a random symmetric state has significantly less entanglement entropy, which is consistent with our analytical investigations.
Panels (\textbf{B},\textbf{C},\textbf{D-I},\textbf{D-II}) show the rescaled purity $R(\ket{\phi})$ of a random state without symmetry, a~random symmetric state, a~random translation invariant state for $\theta =0$, and~that for $\theta = \pi$, respectively. The~rescaled purity is more suitable to see the entanglement phases. The~insets numerically provide $- \ln[ p(R(\ket{\phi} = s)]/2^{2n_A+1}$ as a function of $s$, where $p(R(\ket{\phi} = s)$ is the probability density function. In~the insets, we also plotted quadratic functions (brown dotted lines) fitted to the numerical data as a reference, which may be useful to detect the phase transition. See the main text for the detail.
}
\label{Fig:Distribution}
\end{center}
\end{figure}

In Figure~\ref{Fig:Distribution}, we provide the distributions of entanglement of random states in the subspaces. Panel~(A) is in terms of the entanglement entropy, where we depict the distribution over a Haar random state in the whole Hilbert space $\cH^{\Lambda}$ (red), that in the symmetric subspace $\cH_+^{\Lambda}$ (purple), and~that in the translation invariant subspace $\cH^{\Lambda}_0$ with $\theta = 0$ (blue). As~we showed analytically, the~distribution of random symmetric state significantly differs from the fully random one, whereas that of the random translation invariant state does not. It is however hard to observe any features of entanglement phases since the distribution is highly~concentrated. 

Panels (B), (C), (D-I), and~(D-II) are the distribution of entanglement in terms of the rescaled purity $R$ of a reduced density matrix defined by
\begin{equation}
R(\ket{\phi}) := d^{n_A} \tr[ (\phi^A)^2].
\end{equation}
Note that $R(\ket{\phi}) \in [1, d^{n_A}]$ and is less when the state $\ket{\phi}^{\Lambda}$ is more entangled. Panels (B), (C), (D-I), and~(D-II) are, respectively, for~a random state in the whole Hilbert space, a~random symmetric state, a~random translation invariant state with $\theta = 0$, and~a translation invariant state with $\theta = \pi$.

Let us first check the distribution in the whole Hilbert space (Panel (B)). In~this case, the~probability density function was studied in great detail~\cite{NMV2011}, with~which we compare our numerical result. In~terms of the rescaled purity, it is known that the probability density function $p(R(\ket{\phi} = s)$ over a random state in the whole space has two singularities in the asymptotic limit $n_A \rightarrow \infty$: one is at $s =s_1= 5/4$, and~the other is at $s=s_2 = 2 + 2^{4/3}/2^{n_A/3} \approx 2.79$. These two singularities split the distribution into three entangled phases, namely the maximally entangled phase for $s \in [ 1, s_1)$, the~typical phase for $s \in [s_1, s_2)$, and~the separable phase $s \in [s_2, 2^{n_A}]$.
In our numerics, it is hard to clearly observe the singularities. In~particular, there is no feature of the phase transition at $s=s_1$ at all. This is simply because the probability density function $p(R(\ket{\phi} = s)$ for $s \in [1, s_1)$ scales as $(s-1)^{2^{2n_A-1}}$. Thus, for~$n_A=5$, $p(R(\ket{\phi} = s) = O( (s-1)^{500})$, which is intractable by a numerical sampling method.
On~the other hand, a~trace of the phase transition at $s=s_2$ can be observed from our numerical plot. In~particular, by~looking at the inset of Panel (B), where we plotted $- \ln[ p(R(\ket{\phi} = s)]/2^{2n_A+1}$ as a function of $s$, we observe that the function is quadratic when $s \leq 2$, but~gradually becomes less for $s>2$. This is consistent with the analysis in Ref.~\cite{NMV2011} and can be considered as a feature of the phase transition at $s=s_2$. Note that the phase transition at $s=s_2$ is pointed out to be sensitive to the finite-size effect, resulting in the feature less drastic in our numerics with $n_A$ being $5$.

We now move onto the distributions of the rescaled purity for a random symmetric state and random translation invariant states, which are shown in Panels (C) and (D), respectively. For~random translation invariant states, the~$\theta$ is chosen to be $0$ and $\pi$ in Panels (D-I) and (D-II), respectively, but~similar behaviors are observed for other $\theta$'s.
Although we do not observe clear singularities, which~is similar to the case of a random state in the whole space, the~insets show that
\begin{equation}
- \frac{\ln[ p(R(\ket{\phi} = s)]}{2^{2n_A+1}}
=
\begin{cases}
\text{quadratic in $s$} & \text{for $s \leq s_{\min}$,}\\
\text{linear in $s$} & \text{for $s > s_{\min}$,}
\end{cases}
\end{equation}
where $s_{\min} = {\rm argmin}\bigl[ - \frac{\ln[ p(R(\ket{\phi} = s)]}{2^{2n_A+1}} \bigr]$. This change of the scaling in terms of $s$ shall indicate the presence of the phase transition between the typical and the separable phases. Hence, it seems that even when the state has permutation or translation symmetry, the~typical and the separable phases exist.
On the other hand, it remains open whether the maximally entangled phase exists for random symmetric/translation invariant~states.

\section{Is Generic Entanglement with Symmetry Physical?} \label{S:F}

Before we conclude the paper, we make a remark on the question of whether generic entanglement is physically feasible. In~the case of generic entanglement without symmetry, this question arises from the fact that a Haar random state cannot be efficiently generated by quantum circuits even approximately. Hence, it takes exponentially long time for the distribution of a Haar random state to be achieved by \emph{any} physical dynamics as far as it consists of a-few-body~interactions.

Although it is true that a Haar random state is not physically feasible, recent developments of the theory of \emph{unitary designs}~\cite{DCEL2009} show that the distribution mimicking lower statistical moments of a Haar random state can be quickly generated by quantum circuits~\cite{HL2009,CLLW2015,BHH2016,HM2018,HMHEGR2020} or even by Hamiltonian dynamics~\cite{NHKW2017,OBKBWE2017}. Also, much evidence was obtained that showed that chaotic dynamics result in properties similar to those of a Haar random state~\cite{K2015,K2015Feb,SS2015,RS2015,MSS2016,MS2016}. In~particular, entanglement properties of a Haar random state can be approximately reproduced in many different ways~\cite{ODP2007,DOP2007,NTM2012,NKM2014}.
Thus, generic entanglement, although~it is an idealization in a strict sense, shall be considered to capture characteristic properties of complex quantum many-body systems and hence, physically~feasible.

Regarding the generic entanglement of quantum states with symmetry, an~interesting question from this perspective is that: is it possible to efficiently implement a random state \emph{with symmetry} by quantum circuits or by the dynamics of quantum many-body systems? 
A natural way to achieve this is to first generate a random state by the aforementioned means and then change the basis into symmetric one. It will be however more interesting from the physics perspective if one can find a way that has natural interpretations in terms of Hamiltonian dynamics with reasonably physical Hamiltonian, such~as those with few-body interactions and with less time-dependence.
{To do so in a rigorous manner, it is highly desired to investigate physically feasible constructions of unitary designs with symmetry, which we may call \emph{symmetric} unitary designs.
Since unitary designs transform any pure state to the one that has similar properties of Haar random states, applying a symmetric unitary design to a pure state will reproduce generic entanglement of random states with symmetry that we clarified in this paper. Hence, by~exploiting physically natural constructions of symmetric unitary designs, the~connection of our analysis to complex quantum many-body systems with symmetry will be much more elaborated. Note, however, that a couple of results have been obtained along a similar line~\cite{KVH2018,M2020}, which already indicates that generic entanglement of random states with symmetry reveals characteristic features in those systems.}

\section{Summary and~Discussions} \label{S:Conclusion}
In this paper, we studied how symmetry affects the properties of generic entanglement. Specifically, we  investigated the entanglement entropy of a Haar random state in the invariant subspace with respect to a given symmetry. The~main technical tool is the concentration formula for the entanglement entropy of a random state. We have first extended it to the one applicable for any subspace, and then applied it to invariant subspaces of axial, permutation, and translation symmetries.
It turns out that compared to the entanglement entropy of a random state in the whole Hilbert space, the~axial symmetry often reduces entanglement by a constant degree, and~that there is a significant reduction by the permutation symmetry. In~contrast, the~translation symmetry does not reduce entanglement entropy so much, implying that the same properties of generic entanglement without symmetry shall be observed even in the systems with translation~symmetry.

{Towards the problem of how symmetry affects generic entanglement, these results imply that even when a random state has symmetry, the~concentration formula still holds as shown in Theorem~\ref{Thm:ConEnt}. In~contrast, it is likely that imposing symmetry reduces entanglement, at~least for the symmetries we considered in this paper. The~degree of reduction is, however, highly dependent on what symmetry is imposed. By~closely looking at our results, it is observed that the degree of reduction is related to the size of the invariant subspaces of the symmetry. Whether this is always the case for any symmetry will be left open as a future problem.}

We have also numerically studied the presence of the entanglement phases that are observed for a Haar random state without symmetry. Our numerical analysis is far from conclusive due to the fact that the entanglement phases are sensitive to the finite-size effect, we showed that the typical and separable phases seem to exist even when the state has permutation or the translation~symmetry. 

{We think that our analysis opens a number of open questions. First, as~we mentioned above, it is important to clarify whether or not imposing symmetry always reduces entanglement entropy and, if~so, whether the degree of reduction is always determined by the size of the invariant subspaces. Although~we may naturally expect these to be true, we dealt only with abelian symmetries in this paper. Hence, there still remains a possibility that a random state with non-abelian symmetry may result in more exotic features of entanglement.}

{It will be also interesting to investigate multipartite entanglement of a random state with symmetry. In~the case of a Haar random state without symmetry, this is addressed in Refence~\cite{GFE2009}, where it was shown that most Haar random states are too entangled to be useful as computational resources. Recalling that entanglement is likely to be reduced by imposing symmetry, it may be possible to use random states with symmetry, for~instance the one with permutation symmetry, as~a computational resource. Thus, investigating multipartite entanglement of random states is not only of theoretical interest but may also be of practical use.}

{It is also important to address generic entanglement of mixed states. There are however a number of difficulties around the question. First, unlike the pure state, where a random state can be uniquely defined using the uniqueness of the Haar measure, there is no unique or a priori way to define random mixed states. Although~there are several attempts to define random mixed states, e.g.,~in Ref.~\cite{HLW2006}, it seems that no consensus has been made yet. It is also difficult to evaluate entanglement of mixed states since the entropy of a reduced density matrix is no longer a measure of entanglement. Hence, addressing generic entanglement of mixed states, though~it is an interesting problem, may need more elaborate technique.
}

\section{Acknowledgement}
This research was funded by JST, PRESTO Grant Number JPMJPR1865, Japan.

\end{document}